\documentclass[10pt,onecolumn,a4paper]{article}
\newcommand{\affil}[1]{$^{\rm #1}$}
\usepackage[authoryear]{natbib}
\bibpunct{(}{)}{;}{a}{}{,}
\usepackage{graphicx}
\usepackage{psfig}
\date{} 
\title{Random Numbers from Astronomical Imaging}
\author{{\it Kevin A.\ Pimbblet\affil{A,C} and Michael Bulmer\affil{B}}\\
\affil{A}\,Department of Physics, University of Queensland, Brisbane, Queensland 4072, Australia\\
\affil{B}\,Department of Mathematics, University of Queensland, Brisbane, Queensland 4072, Australia\\
\affil{C}\,E-mail: pimbblet@physics.uq.edu.au}

\begin{document}
 \maketitle
\begin{minipage}{.9\textwidth}
{\bf Abstract}\\
%
This article describes a method to turn astronomical imaging
into a random number generator by using the positions of 
incident cosmic rays and hot pixels to generate bit streams.
We subject the resultant bit streams to
a battery of standard benchmark statistical tests 
for randomness and show that these bit streams
are statistically the same as a perfect random bit stream.
Strategies for improving and building upon this method are outlined.

\medskip{\bf Keywords:}  methods: statistical -- techniques: image processing -- techniques: miscellaneous
\medskip
\end{minipage}

%
%

\section{Introduction}

Random numbers are of importance to many sub-fields of science.
In observational astrophysics they are required for diverse uses
(Meurers 1969) such as testing for sub-structure 
in galaxy clusters (Dressler \& Shectman 1988)
and Monte Carlo background correction techniques (Pimbblet et al.\ 2002).
In cryptography, the generation of secure passwords and cryptographic 
keys is paramount to communication being immune from eavesdropping.
They are also used in selecting winning numbers for lotteries
including the selection of Premium Bonds in the United Kingdom
(http://www.nsandi.com/products/pb/).
Large Monte Carlo computations, however, remain
the primary driver of
intensive searches for truly random number generators
(e.g.\ Ferrenberg, Landau, \& Wong 1992; James 1990).

For many purposes we would essentially like to have a long bit stream
consisting of 1's and 0's.  Each bit in the stream should be independently 
generated with equal probability of being a 1 or 0.  Therefore 
as the length of the stream, $n$, tends toward infinity,
the expectation value of any individual bit being either 1 or 0
is 1/2.  The traditional method of obtaining such a stream is to
use a pseudo-random number generator (PRNG; e.g.\ Press et al.\ 1992).
PRNG's typically rely upon the input of a `seed' quantity which is then
processed using numerical and logical operations to give
a stream of random bits.
Whilst such PRNG's are probably sufficient for most (minor) types 
of applications, they are clearly predictable if the initial 
seed is known.   This makes PRNG's highly inappropriate for
Monte Carlo-like calculations (Gonz{\' a}lez \& Pino 1999).

A truly random number generator (RNG) should possess qualities
that make the bits unpredictable.  The obvious sources of RNG's 
are those that possess large amounts of entropy or chaos (Vavriv 2003;
Gleeson 2002; Gonz{\' a}lez \& Pino 1999).  Examples
include radioactively decaying sources 
(e.g.\ HotBits; http://www.fourmilab.ch/hotbits/),
electrical noise from a semiconductor diode, and thermal noise.

An overlooked and potentially large source of random numbers 
is to be found in astronomical imaging.
Imaging at a telescope will inevitably produce unwanted
cosmetic features such as cosmic ray events, satellite trails
and seeing effects; blurring due to the movement of the atmosphere 
(in the case of ground-based telescopes).
It is precisely these features (and in particular cosmic rays)
which potentially make astronomical imaging a good RNG. 

This article presents an assessment of 
astronomical imaging data as
a source for a RNG.  In Section~2, we demonstrate how it is possible
to generate a stream of random bits from a single astronomical 
image.  Examples of such bit streams are examined in Section~3 using
a battery of statistical tests to evaluate their randomness.
Our findings are summarized in Section~4.

\section{Generating Random Bits}

\begin{figure}
\centerline{
\psfig{file=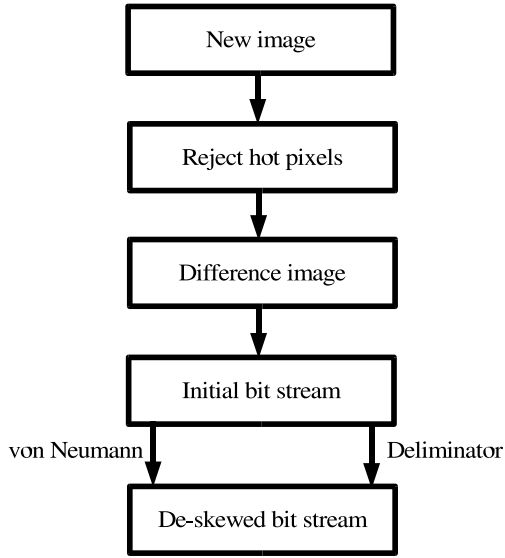,angle=0,width=3.in}
}\vspace*{-1.0in}
\caption{Overview of the processes required to generate
a random bit stream from an initial astronomical image.
On a Dell Precision workstation 530 machine,
analyzing an initial image of $2 \times 4$ k pixels, 
the generation of an initial bit stream 
takes about 50 seconds whilst the de-skewing requires about
30 seconds (for either method).  The average rate of random bit 
production is approximately 2500 bits per second.
Depending upon the amount of discarded bits, this figure can range
from as low as 1000 bits per second up to 4000 bits per 
second.}\label{fig:fchart}
\end{figure}

Assuming that one is in possession of a sample of astronomical
images that possess cosmic ray events we can proceed to obtain
a bit stream from them by following the procedure 
outlined in Figure~\ref{fig:fchart}.  We detail the individual 
steps below.

\begin{figure*}
\centerline{
\psfig{file=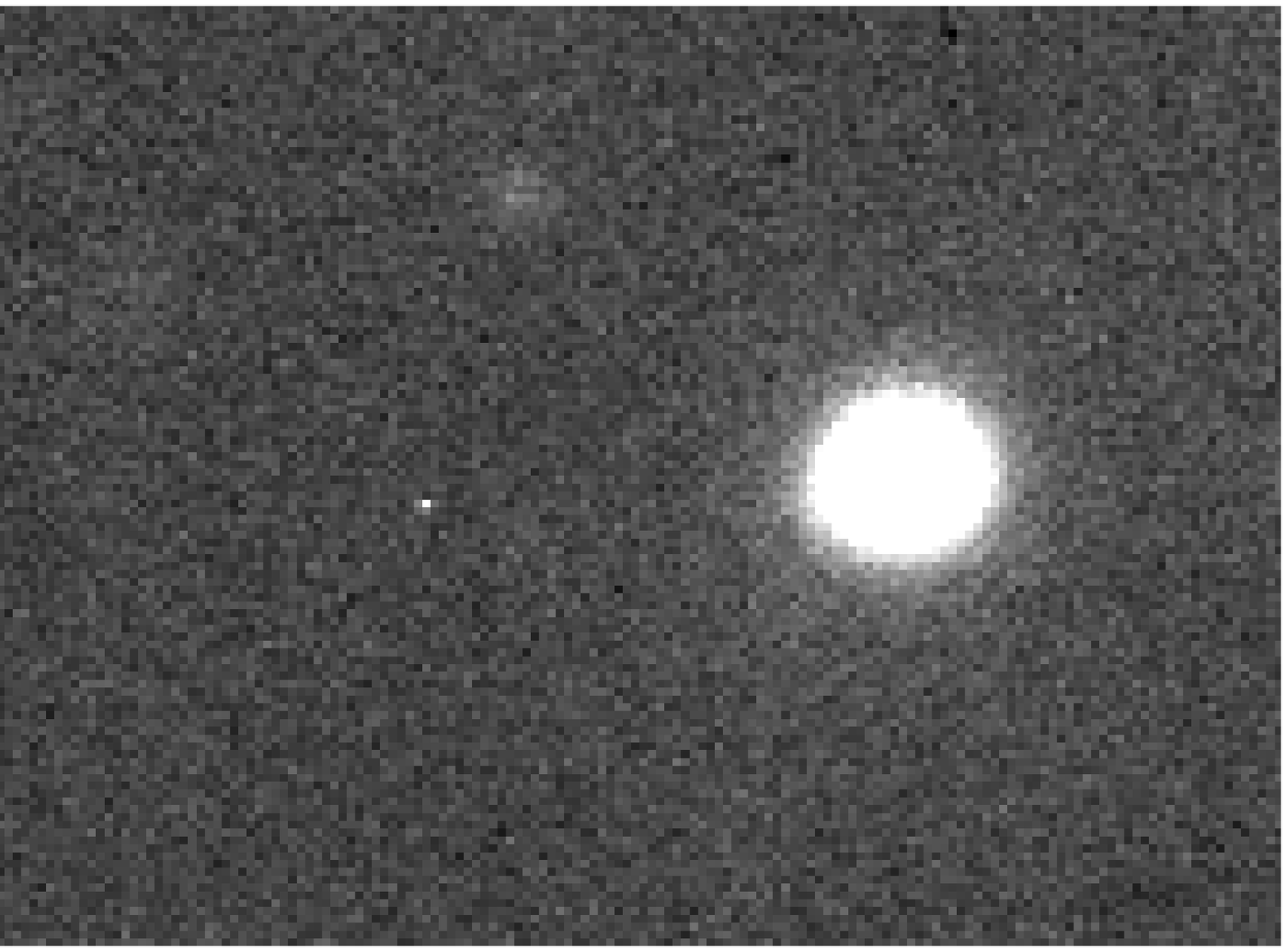,angle=0,width=3.in}
}
\centerline{
\psfig{file=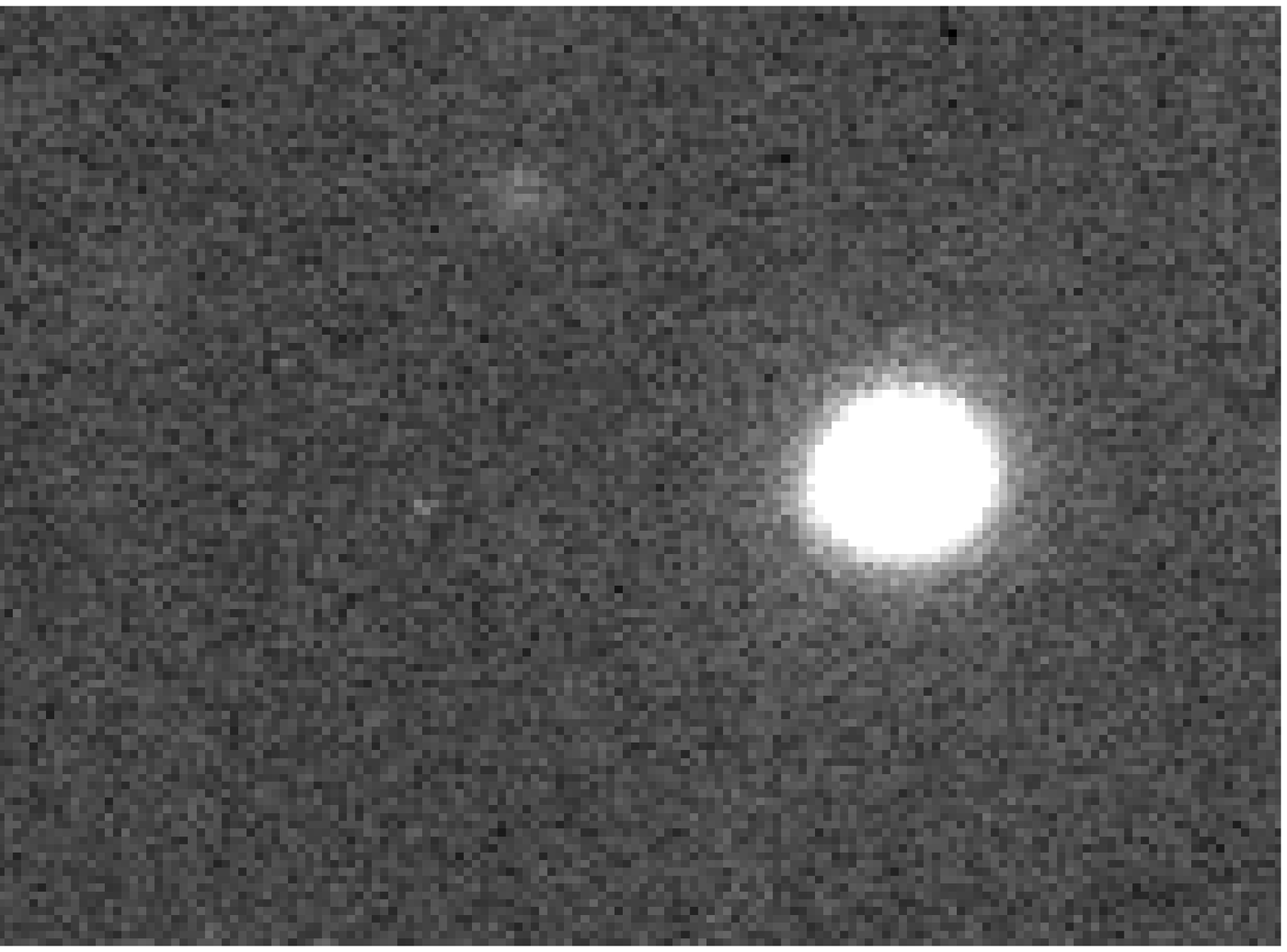,angle=0,width=3.in}
}
\centerline{
\psfig{file=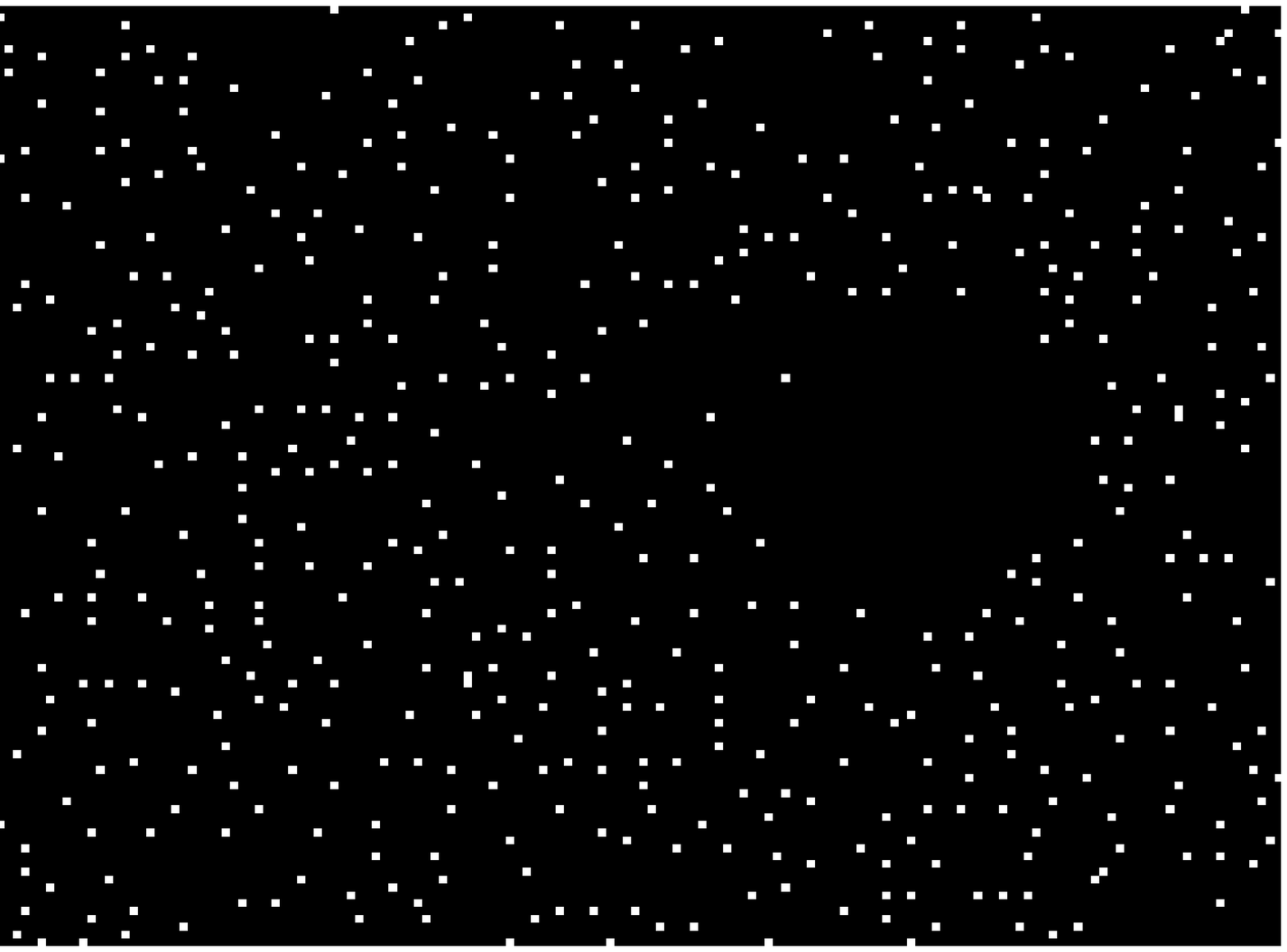,angle=0,width=3.in}
}
\caption{Example of the image processing method.  Top: a sub-section of
the original image measuring $613 \times 480$ pixels.  
Middle: the cosmic ray rejected version
of the image.  Bottom: the difference image.  Note how 
the obvious cosmic ray (left of centre) is rejected along with
a host of other relatively `hot' pixels.  Real objects, meanwhile,
leave an obvious hole in the difference image which requires
de-skewing to generate a random bit stream.}\label{fig:images}
\end{figure*}

Our aim is to detect the locations of any cosmic ray or
`hot spot' (pixel values that are significantly greater than
their local neighbours) in the pixel distribution.  
For this experiment, we use single-shot exposures of 300 to 600 seconds from 
non-overlapping wide-field observations consisting of $2\times4$ k pixels from
Pimbblet and Drinkwater (2004) and their on-going
follow-up observations\footnote{It is also unnecessary to
pre-process these images with flat-fields, for example, as all
we are interested in are the locations of hot pixels.
Indeed, our testing has shown that a raw image produces just an equally 
random bit stream as a post-processed one does.  One problem that 
is encountered is the presence of bad pixels, which always occur in the
same place on a CCD.  These should be removed with the {\sc fixpix} 
(or similar) task before proceeding.}.
Firstly, we use the {\sc IRAF} (http://iraf.noao.edu/)
task {\sc cosmicrays} with default parameters to remove 
the cosmic rays from the original image.  Then, using
{\sc imarith}, we subtract off the cosmic ray free imaging from
the original to create a difference image in which there should
be only cosmic rays (Figure~\ref{fig:images}).  
Inevitably, this technique will identify not only true cosmic rays but
also anomalously hot pixels from the distribution.
To turn the difference image into a bit stream, we sequentially
examine the contents of each pixel in turn, row by row, column
by column.
Pixels with a value of zero in the difference image translate
into a 0 for the bit stream whilst those with values greater
than zero (the hot pixels) become 1.

The fraction of pixels identified as cosmic rays 
(and hot pixels) using this method
is typically 2--3 per cent for our exposures.  
Clearly, there exist more 0's in the bit stream than there are 1's.
Moreover, there are distinct `holes' in the hot pixel distribution
of the difference image where legitimate objects occurred in the
original image (the galaxy in Figure~\ref{fig:images}).  
So, whilst there are random events in our bit stream, it is highly
skewed toward 0's.

\subsection{De-skewing}

To turn our bit stream into a uniformly random distribution,
it is necessary to de-skew it (an `entropy distillation process';
Rukhin et al.\ 2001).  
Here we adopt and investigate 
two common methods of de-skewing.  The first is that of
von Neumann (1963).  We read the bit stream generated from
the imaging as a sequence of non-overlapping pairs.  The
pairs are then transformed into a new bit stream according to
the scheme presented in Table~\ref{tab:vn}.  This scheme 
removes all biases in the original bit stream at the expense of
drastically reducing the overall size of the original as it 
removes the long sequences of 0's associated with legitimate objects
(Table~\ref{tab:compare}).  
The typical reduction for our imaging is in the range 85--98 per cent,
although this is a highly variable parameter.

\begin{table}
\begin{center}
\caption{Scheme for the von Neumann (1963) de-skewing method.
The original bit stream from the imaging is read in as 
a sequence of non-overlapping pairs (`Input Pair' column).  The
output for the new bit stream is then given in 
the `Output' column.  Where `Null' is indicated, nothing
is appended to the new bit stream.}\label{tab:vn}
\begin{tabular}{cc}
\hline Input Pair & Output \\
\hline
00 & Null \\
11 & Null \\
01 & 0 \\
10 & 1 \\
\hline
\end{tabular}
\medskip\\

\end{center}
\end{table}

The second method is to use the hot pixels
(or groups of hot pixels; 1's) as deliminators between long streams of 0's.  
The length of non-overlapping pairs of 
long streams of 0's are then compared to each other to generate a 1
or a 0 depending if the first stream is longer than the second or
vice-versa.
If the lengths are equal, nothing is appended to the new bit stream.
An example of how both of these methods work
is illustrated in Table~\ref{tab:compare}.
The clear disadvantage of the delimination method is that a much
smaller bit stream is produced than for the von Neumann method.

\begin{table*}
\begin{center}
\caption{A comparison of how the de-skewed bit stream
is generated using the von Neumann (1963) and deliminator
methods.}\label{tab:compare}
\begin{tabular}{lc}
\hline  & Bit stream \\
\hline
Pixel Distribution    & \texttt{ 0010000100000010010100100010010001100011010001 }\\
\hline
von Neumann pairings  & \texttt{ ++--++--++--++--++--++--++--++--++--++--++--++ }\\
von Neumann de-skewed & \texttt{ ..1...0.......1.0.0...1...1.0...0.1.....0...0. }\\
\hline
deliminator pairings  & \texttt{ ++++++++----------+++++-------++++++++++------ }\\
deliminator de-skewed & \texttt{ 0.......1.........0....1................0..... }\\
\hline
\end{tabular}
\medskip\\

\end{center}
\end{table*}

\section{Evaluating the Randomness}

In truly random bit stream, each bit should be generated
with probability $1/2$ of producing either a 0 or 1.  
Further, each bit should be generated independently of any
other bit in the bit stream.  One should not, therefore, be able 
to predict the value of a given bit by examining the values
of the bits generated prior to it in the bit stream.
These conditions define an ideal, truly random bit stream
and we will use them to test our random bits against.

To evaluate the randomness of our bit stream, we subject it to a battery
of benchmark statistical tests.  
The tests we use are a selection of those devised by Random Number Generation
and Testing collaboration of the National Institute of Standards and
Technology (NIST; Maryland, USA; http://csrc.nist.gov/rng/index.html).
The NIST statistical test suite source code is freely available
to download from their web site.  

For each test, the software formulates a specific null hypothesis
($H_0$) and alternative hypothesis ($H_1$).  
We will specify that $H_0$ is the hypothesis that our
bit stream is random and that $H_1$ is the hypothesis that
it is non-random.
To accept or reject $H_0$, one determines a test statistic and
compares it to a critical value chosen to be in the tails of
a theoretical reference distribution of the test statistic.  
The possible outcomes of the statistical testing are 
illustrated in Table~\ref{tab:test}.
The probability of obtaining a Type I error (Table~\ref{tab:test})
is therefore the level of significance of a test (see Rukhin et al.\ 2001),
which we set at a level of 0.01 for this work.

\begin{table}
\begin{center}
\caption{Possible configuration of the conclusions 
of any of the given statistical hypothesis tests (Rukhin et al.\ 2001).  
$H_0$ is the hypothesis
that the bit stream is random.}\label{tab:test}
\begin{tabular}{lcc}
\hline True      & \multispan2{\hfil Result \hfil} \\
       Situation & Accept $H_0$ & Reject $H_0$ \\
\hline
$H_0$ true       & Correct  & Type I error \\
$H_1$ true       & Type II error & Correct \\
\hline
\end{tabular}
\medskip\\
\end{center}
\end{table}

The software determines a $P$-$value$ for each test: the probability that
a {\bf perfect} RNG would produce a bit stream that is {\bf less}
random than the bit stream that we test (i.e.\ a $P$-$value$ of zero denotes
a bit stream that is certainly non-random).  
Therefore to reject the null hypothesis, $H_0$, 
(and hence fail the test)
at a 99 per cent confidence level would require: $P$-$value < 0.01$. 
Clearly, the $P$-$value$ only assesses the relative incidence of Type I
errors (Table~\ref{tab:test}).  What it does not describe
is the 
probability that a non-random number 
generator could produce a sequence of numbers at least as random 
as our bit stream that is being tested (a Type II error; 
Table~\ref{tab:test}; Rukhin et al.\ 2001).

Here, we briefly describe each
test (the tests and the statistics behind them are described in 
much more detail in Rukhin et al.\ 2001)
and summarize the results in Table~\ref{tab:res}.

If our bit stream is random, then the number of 1's and 0's
overall and in any part (i.e.\ any sub-sequence) of it should 
be approximately the same.  Therefore the first test is to 
examine the frequency of 1's and 0's in the bit stream.  
In the second test, these frequencies are re-computed for 
sub-sequence blocks of length $M$ (see also Knuth 1981; Pitman 1993).  

Next, we test the number of runs in the sequence, where a run
is defined as an uninterrupted sequence of identical bits.
This tests if the bit stream oscillates with sufficient
celerity between 1's and 0's.  We follow this test by a similar
one that evaluates the longest run of 1's in the sequence to determine
if it is the same as would be expected for a random distribution.
If there is an irregularity in the longest run length of 1's, 
then this will also be reflected in the longest run length of 0's;
hence we only test the longest run length of 1's.

To test for any linear dependence of sub-strings of fixed length
within the original sequence, the rank of disjoint sub-matrices
is examined.  This method is described in more detail by
Marsaglia in the {\sc diehard} statistical 
tests (http://stat.fsu.edu/$\sim$geo/diehard.html).

We can also consider the bit stream as a random walk and hence
test the maximal excursion from zero for the cumulative sum (cusum)
of adjusted digits ($+1, -1$) in our bit stream or sub-sequence 
therein (Revesz 1990).
For a random sequence, this cusum should be near zero.
Finally, by performing a discrete fourier transform (DFT), we can look
for periodic features in our bit stream that would indicate a 
lack of randomness.

\begin{table*}
\begin{center}
\caption{The proportion of 100 bit streams of 
length $n=1,000,000$ that passed each statistical test
(critical $P$-$value=0.01$).  
The minimum 
pass rate in order for our sequence to be considered random
is approximately 0.96 for each statistical test
(see Rukhin et al.\ 2001).}\label{tab:res}
\begin{tabular}{lccl}
\hline Test & \multispan2{\hfil Proportion Passed \hfil} & Notes \\
            & von Neumann & deliminator & \\
\hline
Frequency       & 0.98 & 1.00 & \\
Block Frequency & 0.98 & 1.00 & $M=1000$ \\
Runs            & 0.97 & 0.99 & \\
Longest Run     & 0.99 & 0.97 & \\
Rank            & 1.00 & 0.99 & \\
Cusum           & 0.97 & 0.99 & \\
DFT             & 1.00 & 1.00 & \\
\hline
\end{tabular}
\medskip\\

\end{center}
\end{table*}

\begin{figure*}
\centerline{
\psfig{file=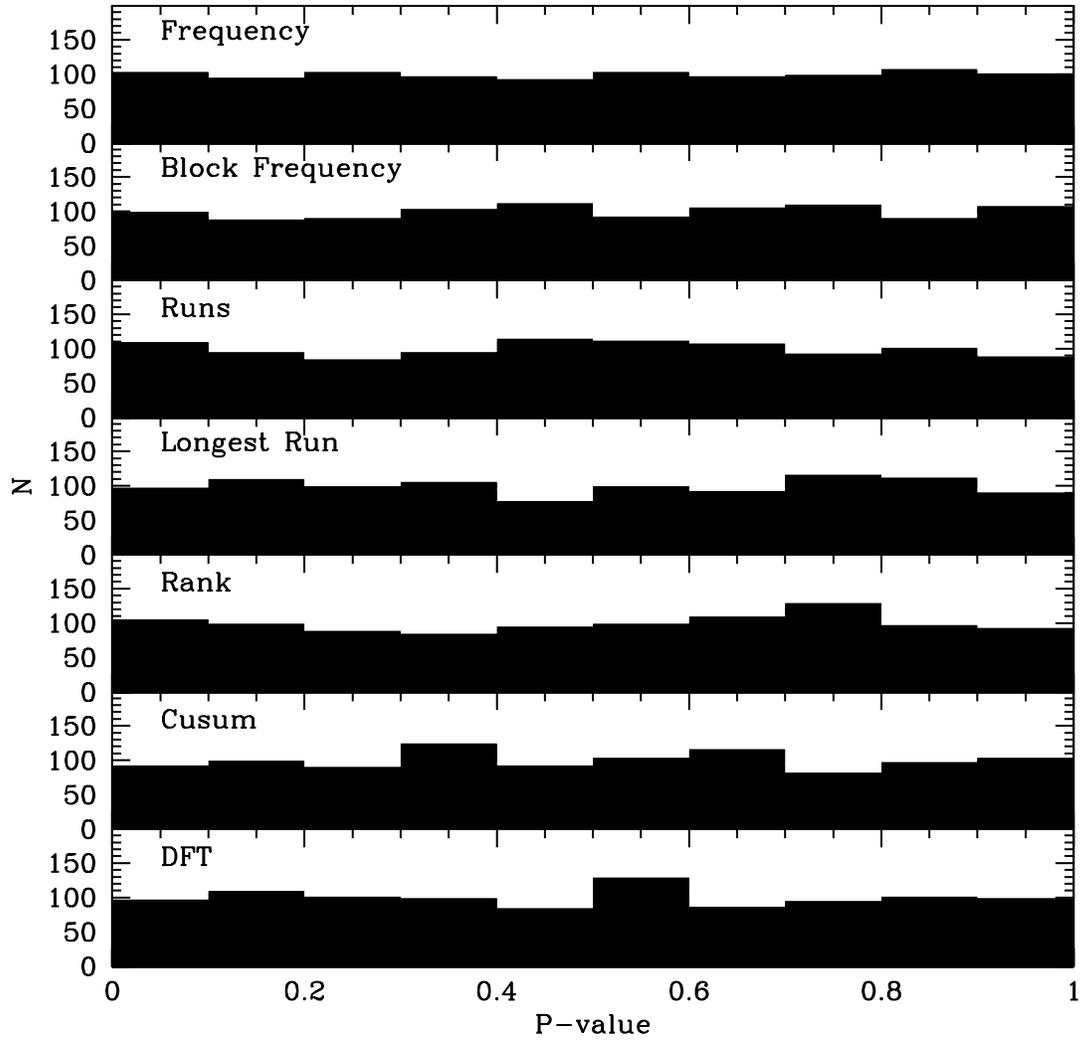,angle=0,width=6in}
}
\caption{Histograms of the $P$-$value$ distributions
arising from applying the seven statistical tests 
from Table~\ref{tab:res} to 1000 bit streams of length
100,000.  
All of the distributions are approximately
uniform.}\label{fig:hist}
\end{figure*}

Table~\ref{tab:res} shows that both variants of the de-skewing
method are sufficient to pass all of the standard tests outlined
above by at least the minimum pass rate (Rukhin et al.\ 2001).  
We can further assess the validity of our conclusion by 
examining the distribution of $P$-$value$s, which for a random
sequence should be $\sim$ uniform.
For this, we re-run our experiment, but use 1000 bit streams
of length 100,000 (since 100
bit streams comprise a relatively small sample).  
All the tests outlined above (Table~\ref{tab:res}) are
passed once again and we display the distribution of
$P$-$value$s for these in Figure~\ref{fig:hist}.
All of the distributions of $P$-$value$s approximate uniformity 
very well.  
To test the uniformity of the distributions, we can use
a $\chi^2$ test and a determination of a $P$-$value$ 
that corresponds to the goodness-of-fit distributional test
of the $P$-$value$s
(a so-called `$P$-$value$ of $P$-$value$s'; Rukhin et al.\ 2001).
The $\chi^2$ statistic is simply:
\begin{equation}
\chi^2 = \sum^{10}_{i=1}\frac{(F_i-100)^2}{100}
\end{equation}
where $F_i$ is the number of $P$-$value$s in bin $i$
of Figure~\ref{fig:hist}.  The $P$-$value$ of $P$-$value$s is then
the complemented incomplete gamma function:
\begin{equation}
1-\Gamma(a,z)
\end{equation}
where we set $a=9/2$ and $z=\chi^2/2$ (see Rukhin et al.\ 2001).
This yields a mean $\chi^2$ value for the distributions
in Figure~\ref{fig:hist} of $9.9\pm2.8$ whilst
$1-\Gamma(9/2,\chi^2/2)=0.357$.  Since $1-\Gamma(9/2,\chi^2/2)$
is much larger than (say) 0.0001, we can consider 
the distributions to be uniformly spread.

We note, however, that by altering the
size of the bit stream downward (say $n=10,000$), we have been able to
cause the von Neumann de-skewed variant to fail the runs test.
This emphasizes the fact that these tests need to be carried out
on a large bit stream sample (at least $n\geq100,000$).

\section{Summary}

We have described how astronomical imaging can be used as a
true RNG by application of simple cosmic ray rejection algorithms.
Although we throw away a large fraction of our original data through
the de-skewing methods, we have shown that resultant 
bit stream is sufficiently random to pass modern tests 
for randomness.  

The tests that we applied are only a selection of 
the NIST statistical test suite.  There are more within this suite
and certainly more beyond (e.g.\ Tu \& Fischbach 2003; Ballesteros \& 
Mart{\'{\i}}n-Mayor 1998; Knuth 1981).
We have therefore looked at applying more complex tests for randomness,
as detailed in the NIST test suite (e.g.\  non-overlapping
template matching, etc.).  
We find that these additional tests are readily passed by both de-skewed
variants of our bit streams.

Several improvements to our methodology can potentially be made.  
We are looking at different de-skewing techniques to 
improve the test statistics.  For example, the von Neumann
method can be used twice (or more) on the bit stream generated from
the image.  The resulting proportions of bit streams that pass the
statistical tests in Table~\ref{tab:res} increases fractionally as
a result, but not significantly.
Our next step is to attempt to create a web interface where it will
be possible to download random numbers in real time 
using this method.  This could be accomplished by using the
international network of continuous cameras (concams; Nemiroff \& 
Rafert 1999).  The concams have the virtues that one does not require
the sky to be dark locally and 
the images are freely available to the public.

The imaging used in this work is at optical wavelengths (specifically
$B$, $V$, $R$ and $I$-bands).  It may be interesting to examine 
how the test results varied with other parts of the electromagnetic
spectrum, if at all.  The imaging is also non-overlapping.  If the 
concams constitute a valid RNG, then it is worthwhile to confirm
that images of the same area of sky produce independent 
random bit streams, which should be the case as we are only
considering the incidence of hot pixels (i.e.\ cosmic rays).

\section*{Accessory Materials}
One sequence of 1,000,000 bits (von Neumann de-skewed; approximately 1.1 Mb
in size) used in this work will be presented in the
accessory materials available from PASA online.
{\it Please note that this bit stream is only one small part of the
much larger sample used to generate the results presented in this
work.}  

\section*{Acknowledgments} 
This work has made use of the NIST statistical test suite.  We
wish to warmly thank NIST for allowing public access to this code.
We also thank the anonymous referee for a very positive and
thorough review that has improved the clarity of this work.
KAP acknowledges an EPSA University of Queensland Fellowship.
The imaging used in this work is from the WFI instrument used
on both the Anglo-Australian Telescope and Siding Spring Observatory
40'' Telescope and we thank both observatories for the time
allocated to us.


\end{document}